\documentclass[12pt]{iopart}

\usepackage{iopams}

\input epsf.sty

\newlength{\TZ}
\newcommand{\BEQ}{\begin{equation}}     
\newcommand{\BEA}{\begin{eqnarray}}
\newcommand{\EEQ}{\end{equation}}       
\newcommand{\EEA}{\end{eqnarray}}
\newcommand{\eps}{\varepsilon}          
\newcommand{\D}{{\rm d}}                
\newcommand{\II}{{\rm i}}               
\newcommand{\wit}[1]{\widetilde{#1}}    
\newcommand{\wht}[1]{\widehat{#1}}      

\renewcommand{\vec}[1]{\boldsymbol{#1}} 

\newcommand{\annexe}[1]{\setcounter{equation}{0}\setcounter{subsection}{0}
\section*{Appendix. #1}
\renewcommand{\theequation}{A\arabic{equation}}
              \renewcommand{\thesection}{A} }

\catcode`\@=11
\def\numberbysection{\@addtoreset{equation}{section}
        \def\theequation{\thesection.\arabic{equation}}}
\numberbysection

\begin{document}

\title{Kinetics of the long-range spherical model}

\author{Florian Baumann$^{a,b}$, Sreedhar B. Dutta$^{c}$ and 
Malte Henkel$^a$} 
\address{$^a$Laboratoire de Physique des 
Mat\'eriaux,\footnote{Laboratoire associ\'e au CNRS UMR 7556} 
Universit\'e Henri Poincar\'e Nancy I, \\ 
B.P. 239, F -- 54506 Vand{\oe}uvre l\`es Nancy Cedex, France}
\address{$^b$Institut f\"ur Theoretische Physik I, 
Universit\"at Erlangen-N\"urnberg, \\
Staudtstra{\ss}e 7B3, D -- 91058 Erlangen, Germany}
\address{$^c$School of Physics, Korea Institute for Advanced Study, \\
Cheongnyangni 2-dong, Dongdaemun-gu, Seoul 130-722, South Korea}


\begin{abstract}
{The kinetic spherical model with long-range interactions is
studied after a quench to $T < T_c$ or to $T = T_c$. For the
two-time response and correlation functions of the
order-parameter as well as for composite fields such as the
energy density, the ageing exponents and the corresponding
scaling functions are derived. The results are compared to
the predictions which follow from local scale-invariance.}
\end{abstract}
\pacs{05.40.-a, 05.10.Gg, 05.70.Ln, 64.60.Ht, 81.15.Aa,
05.70.Np}
\maketitle

\setcounter{footnote}{1}

\section{Introduction}

A many-body system rapidly brought out of some initial state by
quenching it to either a critical point or into a coexistence  region of the
phase-diagram where there are at least two equivalent equilibrium states
undergoes {\em ageing} \cite{Bray94}. 
For ageing systems the physical state evolves slowly, 
non-exponentially and depends on the time since
the quench was performed and hence time-translation invariance 
is broken. In addition, there holds some kind of dynamical scaling,
whether or not the stationary states are critical.
These aspects of ageing can be 
conveniently studied through the two-time response and correlation 
functions defined as
\BEQ
\label{scaling_r}
R(t,s) = \left.\frac{\delta\langle\mathcal{O}(t,\vec{r})\rangle}{\delta 
h(s,\vec{r})}
\right|_{h=0} = s^{-a-1} f_R\left(\frac{t}{s}\right) \;\;,\;\;
f_R(y) \stackrel{y \rightarrow \infty}{\sim}
y^{-{\lambda_R}/{z}} ,
\EEQ
\BEQ
\label{scaling_c}
C(t,s) = \bigl\langle \mathcal{O}(t,\vec{r})\mathcal{O}(s,\vec{r})\bigr\rangle 
= s^{-b} f_C\left(\frac{t}{s}\right) \;\;,\;\;
f_C(y) \stackrel{y \rightarrow \infty}{\sim} y^{-{\lambda_C}/{z}},
\EEQ
where the observable $\mathcal{O}(t,\vec{r})$ (at time $t$
and location $\vec{r}$) is
typically taken to be the order-parameter $\phi(t,\vec{r})$. In this work,
we shall also study composite fields such as the 
energy density. We denote by $h$ the field conjugate to $\mathcal{O}$
(and when $\cal O$ is the order-parameter the conjugate field $h$ is the
associated magnetic field). 
The dynamical scaling forms (\ref{scaling_r},\ref{scaling_c}) 
are expected to hold in the scaling limit where both $t,s\gg t_{\rm micro}$ 
and also $t-s\gg t_{\rm micro}$,
where $t_{\rm micro}$ is some microscopic time scale. In writing
eqs.~(\ref{scaling_r},\ref{scaling_c}), it is implicitly assumed that 
the underlying dynamics is such that there is a single relevant 
length-scale $L=L(t)\sim t^{1/z}$, where $z$ is the dynamical 
exponent (the presence of another relevant large length scale 
would break dynamical scaling). 
Non-equilibrium universality classes are distinguished by different
values of exponents such as $a,b,\lambda_C,\lambda_R$ (which will depend on 
the observable $\cal O$ and the field $h$ used and also on whether $T<T_c$ or
$T=T_c$). 
For reviews, see  \cite{Bray94,Godreche02,Calabrese05,Cugl02,Lux}. 

In trying to find a systematic approach to determine the scaling
functions $f_{R,C}$ it has been proposed to generalise the dynamical scaling to
a {\em local} scale-invariance \cite{Henkel94,Henkel02} which include the
transformation $t\mapsto (\alpha t+\beta)/(\gamma t+\delta)$ in time with
$\alpha\delta-\beta\gamma=1$. In a field-theoretical setting 
\cite{Taeuber,Taeuber07} the autoresponse function can be
formally rewritten as a correlator $R(t,s) = \langle 
\phi(t)\wit{\phi}(s)\rangle$ where $\wit{\phi}$ 
is the response field associated to
$\phi$. From the assumption that both $\phi$ and $\wit{\phi}$ are so-called
{\em quasi-primary} scaling operators
\cite{Bela84,Henkel02} (see section 3), it follows
that
\BEQ \label{1:fR}
f_R(y) = f_0 y^{1+a'-\lambda_R/z} (y-1)^{-1-a'} \Theta(y-1) ,
\EEQ
where $\Theta(y)$ is the Heaviside function which expresses causality, $a'$
is an exponent and $f_0$ a normalisation constant. A similar explicit, if
lengthy, expression can be given for the autocorrelation. We refer to 
\cite{Henkel07a,Henkel07b,Henkel07c} for recent reviews on the 
derivation of these results
and on the numerous examples where these predictions have
been tested. 
For our purposes, it is enough to note that most of these tests are done in 
situations where the dynamical exponent $z=2$. In
particular, almost all existing
tests for $z\ne 2$ merely tested the prediction (\ref{1:fR}), and this for
the order-parameter {\em only}, see \cite{Henkel07a,Henkel07b} for a
detailed discussion. The only exception are a few
simple models where $z = 4$ \cite{Roethlein06,Baumann07b}.

A fuller picture on the validity of the several technical assumptions which
are needed for the precise formulation of the theory of
local scale-invariance (LSI)
can only come from more systematic tests of  its predictions. To this end, 
we shall study in this paper the ageing behaviour of the spherical model with
long-range interactions. It was shown by Cannas, Stariolo
and Tamarit \cite{Cann01} 
that for quenches to $T<T_c$, if the exchange couplings decay sufficiently 
slowly with the distance then the dynamical exponent $z$ becomes a continuous 
function of the control parameters of the model and that the scaling forms
(\ref{scaling_r},\ref{scaling_c}) hold for the order-parameter. Here we
shall extend these
considerations to the critical case $T=T_c$ and shall further look at
the scaling behaviour of composite operators (i.e. energy density). 
Specifically, we shall inquire
\begin{enumerate}
\item whether dynamical scaling holds, and if so, what are the values of the
corresponding non-equilibrium exponents  ?
\item what is the form of the scaling functions of responses  and correlators ? 
\item which of the composite operators, if any, transform as quasi-primary
fields under local scale-invariance ?
\end{enumerate}

In section~2, we review the exact solution of the kinetic long-range
spherical model and list our results for the non-equilibrium exponents and the
scaling functions for the order-parameter and for composite
fields. Some of the details are treated in the appendix. In section~3,
we first show that the presently available formulation \cite{Henkel02} 
of local scale-invariance cannot explain our results on the space-time 
form of the response functions when $z\ne 2$.
We then announce some results of a forthcoming paper \cite{Baumann07} 
on a general reformulation 
of local scale-invariance for $z\ne 2$ before comparing our explicit results
with the corresponding
predictions of that general theory. In section~4 we conclude. 

\section{Exact solution of the long-range spherical model}
\label{exact-sol}

The two-time correlation- and response-functions of the order-parameter in the 
spherical model when quenched either to $T=T_c$ or else to $T<T_c$
are well-known in the case of nearest-neighbour interactions
\cite{Jans89,Newm90,Godreche00,Corberi02,Annibale06}.
These are also known for the long-range model 
when quenched to $T<T_c$ \cite{Cann01}. Here, we shall derive the response and
correlation functions of the order-parameter and of certain composite operators
in the long-range mean spherical model quenched to $T\leq T_c$. 

\subsection{Long-range spherical model}
\label{LRSM}

The long-range spherical model is defined in terms of a real spin variable 
$S(t,\vec{x})$ at time $t$ and on the sites $\vec{x}$ of a $d$-dimensional
hypercubic lattice $\Lambda\subset\mathbb{Z}^d$, subject to the (mean) 
spherical constraint
\BEQ
\label{constraint}
\left\langle \sum_{\vec{x}\in\Lambda} S(t,\vec{x})^2 \right\rangle 
= {\cal N} ,
\EEQ
where $\cal N$ is the number of sites of the lattice\footnote{For short-ranged
interactions, a careful analysis \cite{Fusco02} has shown that the long-time
behaviour is not affected whether (\ref{constraint}) is assumed exactly or
on average.}. The Hamiltonian is given by \cite{Joyce66}
\BEQ
\label{hamiltonian}
{\cal H} = -\frac{1}{2}\sum_{\vec{x},\vec{y}} J(\vec{x}-\vec{y})S_{\vec{x}} 
\left( S_{\vec{y}} - S_{\vec{x}}\right),
\EEQ
where the sum extends over all pairs ($\vec{x},\vec{y}$) such that
$\vec{x}-\vec{y}\ne \vec{0}$. The coupling constant $J(\vec{x})$ of the model is defined by
\BEQ
\label{coup}
J(\vec{x}) = 
\left({\sum_{\vec{y}\in \Lambda}}'|\vec{y}|^{-(d+\sigma)} \right)^{-1}\: 
|\vec{x}|^{-(d+\sigma)} ,
\EEQ 
when $\vec{x} \ne 0$ and vanishes when $\vec{x} = 0$; the summation is over 
all lattice sites except $\vec{y}= 0$. The last term in (\ref{hamiltonian}), 
$\sum_{\vec{x},\vec{y}} J(\vec{x}-\vec{y})S^2_{\vec{x}}$, can also be absorbed
into the Lagrange multiplier that imposes the spherical constraint, see below.

The `usual' spherical model with short-range interactions is given by
$J_{sr}(\vec{x}-\vec{y})= 
J \sum_{\vec{\mu}(\vec{x})} \delta_{\vec{y},\vec{x}+\vec{\mu}(\vec{x})}$, where 
$\vec{x}+\vec{\mu}(\vec{x})$ runs over all the neighbouring sites of $\vec{x}$. When 
$\sigma \ge 2$, the relevant large-scale behaviour of the above model, 
(\ref{hamiltonian}) and (\ref{coup}), is governed by this short-range 
model. Here we shall focus on truly long-range interactions such that $0<\sigma<2$. 
In this case, the dynamical exponent $z=\sigma$ can be 
continuously varied by tuning this parameter, see \cite{Joyce66,Cann01} and below. For the equilibrium behaviour of the model, consult the classic review
by Joyce \cite{Joyce66}. 

The dynamics is governed by the Langevin
equation\footnote{In eq. (\ref{langevin}), fluctuations in
the Lagrange multiplier $\mathfrak{z}(t)$ are neglected. As
pointed out in \cite{Annibale06}, these must be taken into
account when treating non-local observables involving spins
from the entire lattice or if the initial magnetisation is
nonzero. Here we are only interested 
in local quantities and use a vanishing initial
magnetisation. See \cite{Baldovin07} for a 
careful discussion on the applicability
of Langevin equations in long-ranged systems.}  
\BEQ
\label{langevin}
\partial_t S(t,\vec{x}) =  
- \left. \frac {\delta {\cal H} } {\delta S_{\vec{x}} } \right|_{S_{\vec{x}} 
\rightarrow S(t,\vec{x}) }
- \mathfrak{z}(t) S(t,\vec{x}) + \eta(t,\vec{x}) ,
\EEQ
where the coupling to the heat bath at temperature $T$ 
is described by a Gaussian noise $\eta$ of vanishing average and a variance
\BEQ
\label{noise-corr:R}
\left\langle \eta(t,\vec{x})\eta(t',\vec{x}')\right\rangle = 
2T \delta(t-t')\delta(\vec{x}-\vec{x}'). 
\EEQ
The Lagrange multiplier $\mathfrak{z}(t)$ is fixed by the mean spherical constraint. 

The Langevin equation and the variance of the noise in the Fourier space read 
\BEQ
\label{eqn-of-motion}
\partial_t \wht{S}(t,\vec{k}) =  
- \Big( \omega(\vec{k}) + \mathfrak{z}(t) \Big) \wht{S}(t,\vec{k}) + 
\wht{\eta}(t,\vec{k}) ,
\EEQ
\BEQ
\label{noise-corr}
\left\langle \wht{\eta}(t,\vec{k})\wht{\eta}(t',\vec{k}')\right\rangle = 
2 T (2\pi)^d  \delta(t-t') \delta(\vec{k}+\vec{k}'),
\EEQ
where
$\omega(\vec{k}) = \wht{J}(\vec{0}) - \wht{J}(\vec{k})$. The hatted functions denote the
Fourier transform of the corresponding functions. In the long-wavelength limit 
$|\vec{k}| \rightarrow 0 $, the function $\omega(\vec{k}) \rightarrow 
B |\vec{k}|^{\sigma}$, where the constant $B$ is given by \cite{Cann01} 
$B = \lim_{|\vec{k}| \rightarrow 0}\,
(\wht{J}(\vec{0}) - \wht{J}(\vec{k})) |\vec{k}|^{-\sigma}$.

The solution of the above equation is
\BEQ
\label{solution}
\wht{S}(t,\vec{k}) =  \frac {e^{-\omega(\vec{k}) t} } {\sqrt{g(t;T)}}
\left[ \wht{S}(0,\vec{k}) +
\int_0^t \!\!\D\tau \, e^ {\omega(\vec{k}) \tau } \sqrt{g(\tau;T)}
\wht{\eta}(\tau,\vec{k}) \right] ,
\EEQ
with the constraint function $g(t;T) = \exp(2\int_0^t \!\D\tau\, \mathfrak{z}(\tau))$. 
The system is assumed to be quenched from far above the critical temperature, 
hence $\langle \wht{S}(0,\vec{k})\rangle =0$;
and the spins are assumed to be uncorrelated initially, hence the spherical 
constraint implies
$\langle \wht{S}(0,\vec{k})\wht{S}(0,\vec{k}')\rangle 
= (2\pi)^d \delta(\vec{k} +\vec{k}')$.
Therefore, the spin-spin correlation function when $t > s$ is 
\BEQ
\label{corr-fn:k}
\langle \wht{S}(t,\vec{k})\wht{S}(s,\vec{k}')\rangle 
= (2\pi)^d \delta(\vec{k} +\vec{k}') \wht{C}(t,s;\vec{k}),
\EEQ
where
\BEQ
\label{C-fn:k}
\wht{C}(t,s;\vec{k}) = 
\frac {e^{-\omega(\vec{k}) (t + s)} } {\sqrt{g(t; T)g(s;T)}}
\left[ 1 + 2 T
\int_0^{s} \!\!\!\D\tau \, e^ {2 \omega(\vec{k}) \tau } g(\tau;T) \right]. 
\EEQ
The spherical constraint implies $1 = \int_{\Lambda_{\vec{k}}} \wht{C}(t,t; \vec{k})$ and 
gives $g(t;T)$ as the solution to the Volterra 
integral equation \cite{Godreche00,Cann01}
\BEQ
\label{volt}
g(t;T) = f(t) + 2T \int_0^t\!\D\tau f(t-\tau) g(\tau;T) ,
\EEQ
with $g(0;T) = 1$, and $f(t) = f(t,\vec{0})$ is obtained from the function 
\BEQ
\label{f-fn-def}
f(t;\vec{r}) := \int_{\Lambda_{\vec{k}}} \!\D \vec{k}\:
\exp\left( \II \vec{k} \cdot \vec{r} -2\omega(\vec{k})t\right) ,
\EEQ
where $\Lambda_{\vec{k}}$ denotes the first Brillouin zone
of the lattice $\Lambda$. 

\subsection{Composite operators: Correlations and responses}
\label{corr-resp}
We shall now consider not only the spin operator $S(t,\vec{r})$ but also some 
composite fields, specifically the spin-squared (spin$^2$) operator and the 
energy-density operator. We denote the spin and spin$^2$ operators by 
\BEA
{\cal O}_1(t,\vec{x}) &:=& S(t,\vec{x}), \\ 
{\cal O}_2(t,\vec{x}) &:=& S^2(t,\vec{x}) - 
\langle S^2(t,\vec{x}) \rangle, 
\EEA
respectively. 
The energy-density operator is defined as 
\BEA
\label{observ}
{\cal O}_{\epsilon}(t,\vec{x}) := {\cal E}(t,\vec{x}) - 
\langle {\cal E}(t,\vec{x}) \rangle ~~~, \nonumber \\
{\cal E}(t,\vec{x}) := \sum_{\vec{x}'} J(\vec{x}-\vec{x}')
S(t,\vec{x}) \left( S(t,\vec{x}') - S(t,\vec{x})\right).
\EEA
These composite operators are defined in such a way that their average value is 
zero, and hence their correlation functions are essentially the {\it connected} 
correlation-functions. Also note that since the energy is defined only up to a 
constant there is no unique definition of the energy-density operator. 

The distinction between ${\cal O}_2$ and $\mathcal{O}_{\epsilon}$ 
might be better understood
as follows. We look into the 
continuum limit of the energy-density
operator, at least for short-range model, for we shall later discuss that   
this operator is not quasi-primary under local scale-invariance.
In the short-range model, the expression for energy in lattice models 
is usually taken as  
\begin{equation}
\label{energy}
{\cal H} = - J \sum_{\vec{x},\vec{\mu}(\vec{x})} S_{\vec{x}}
S_{\vec{x}+\vec{\mu}(\vec{x})},
\end{equation}
where $\vec{x}+\vec{\mu}(\vec{x})$ runs over the neighbouring sites of $\vec{x}$. 
In such a case, the energy density could be defined as
$\widetilde{\epsilon}(\vec{x}) = - J \sum_{\mu}
S_{\vec{x}} S_{\vec{x}+\vec{\mu}}$, which in the continuum limit would reduce to
$\widetilde{\epsilon}(\vec{x})
= - J \left( 2 S_{\vec{x}}^2 + \mu^2 S_{\vec{x}} \nabla^2
S_{\vec{x}}  \right)$, where $\mu$ is the lattice constant.
But if we had added an overall constant $E_0 = {\cal N} = \sum_x S_x^2$ then the
energy density could be defined as
\begin{equation}
\epsilon(\vec{x})  = - J \sum_{\vec{\mu}} S_{\vec{x}} \left( S_{\vec{x}+\vec{\mu}} -
S_{\vec{x}} \right) \rightarrow
- J \mu^2 S_{\vec{x}} \nabla^2 S_{\vec{x}} 
\left( 1 + {\rm O}(\mu)\right).
\end{equation}
Hence ${\cal H}_{\rm sr} = \sum_{\vec{x}} \epsilon(\vec{x})  
= J \mu^2 \sum_{\vec{x}} (\nabla S_{\vec{x}})^2$, up to boundary terms. Therefore, for our 
model (\ref{hamiltonian}) the two operators 
${\cal O}_2(t,\vec{x})$ and ${\cal O}_{\epsilon}(t,\vec{x})$ must be  
distinguished.

The connected two-point correlation functions of the composite operators 
\BEQ
\label{corr-def}
{\cal C}_{ab}(t,s;\vec{x} - \vec{x}') := 
\langle {\cal O}_a(t;\vec{x}) {\cal O}_b(s;\vec{x}')\rangle
\EEQ 
are obtained by making use of Wick's contraction as 
detailed in the appendix. Throughout it is implicitly assumed that 
$t > s$ unless stated otherwise. As we have spatial-translation invariance 
in our system, we shall find that all two-point quantities depend
merely on the difference $\vec{r} := \vec{x}-\vec{x}'$ of
the spatial coordinates.

The response functions of the fields $\{ {\cal O}_a(t,\vec{x}) \}$
to the conjugate fields $\{ h_a(t,\vec{x})\}$
\BEQ
\label{resp-def}
{\cal R}_{ab}(t,s,\vec{x} - \vec{x}') := \left.
\frac{\delta \langle {\cal O}_a(t,\vec{x}) \rangle_{\{h\}} }
{\delta h_b(s,\vec{x}')} 
\right|_{\{h\}=\{0\}},
\EEQ 
are obtained by linearly perturbing the Hamiltonion, ${\cal H} \rightarrow 
{\cal H} - \sum_{a,t,\vec{x}} h_a(t,\vec{x}) {\cal
O}_a(t,\vec{x})$, as detailed in 
the appendix. The above defined response function can be interpreted as the
susceptibility of the expectation value of a field to 
near-equilibrium fluctuations.

Finally, we also obtain out-of-equilibrium responses of the fields 
$\{ {\cal O}_a(t,\vec{x}) \}$ to local temperature
fluctuations. This we do by perturbing the 
noise strength $T \rightarrow T +  \delta T(t,\vec{x})$ and then evaluating 
the response functions
\BEQ
\label{temp-resp-def}
{\cal R}^{(T)}_{a}(t,s,;\vec{x} - \vec{x}') := \left.
\frac{\delta \langle {\cal O}_a(t,\vec{x}) \rangle_{\delta T} }
{\delta T(s,\vec{x}')} 
\right|_{\delta T=0}.
\EEQ 
Let us specify at this point the asymptotic scaling
forms that we expect for the autocorrelation function
${\cal C}_{ab}(t,s) := {\cal C}_{ab}(t,s; \vec{0})$ and the
autoresponse functions ${\cal R}_{ab}(t,s) := {\cal R}_{ab}(t,s;\vec{0})$ 
and ${\cal R}^{(T)}_{a}(t,s) := {\cal R}^{(T)}_{a}(t,s;\vec{0})$. 
They are expected to behave as
\BEA
\label{expC}
{\cal C}_{ij}(t,s) = s^{-b_{ij}} f^{ij}_{C}(t/s),&\qquad&
f^{ij}_C(y) \stackrel{y \rightarrow
\infty}{\sim}y^{-\lambda^{ij}_{C}/z}, \\ 
\label{expR}
{\cal R}_{ij}(t,s) = s^{-a_{ij} -1} f_R^{ij}(t/s),&\qquad&
f^{ij}_{R}(y) \stackrel{y \rightarrow \infty}{\sim}
y^{-\lambda^{ij}_{R}/z}, \\
\label{expRT}
{\cal R}_{i}^{(T)}(t,s) = s^{-a_{i}^{(T)} -1} f_R^{(T)i}(t/s),&\qquad&
f_R^{(T)i}(y) \stackrel{y \rightarrow \infty}{\sim}
y^{-\lambda^{(T)i}_{R}/z}, 
\EEA
in the scaling regime where $t,s$ and $t-s$ are
simultaneously large. This also defines the nonequilibrium
critical exponents $a_{ij},b_{ij}, a_i^{T},
\lambda^{ij}_{R},\lambda^{ij}_{C}, \lambda_{R}^{(T)i}$.

We now write the correlation and response functions of some of the fields
$\{ {\cal O}_a(t,\vec{x}) \}$ in terms of the spin-spin correlator $C(t,s;\vec{r})$, 
the constraint function $g(t; T)$ and $f(t;\vec{r})$. 
The details of these computations are given in the appendix, while the explicit forms of 
these functions and their asymptotics are spelt out in the
next subsection.  

\subsubsection{The correlation functions:}~\\
We obtain the following expressions for the non-vanishing 
correlation functions of the 
composite fields.
\begin{itemize}
\item
The spin$^2$-spin$^2$ correlation function is found to be
\BEQ
\label{s2-s2-corr}
{\cal C}_{22}(t,s;\vec{\vec{r}}) = \langle
\mathcal{O}_2(t,\vec{r}) \mathcal{O}_2(s,\vec{0}) \rangle =
2 \Big[ C(t,s;\vec{r}) \Big]^2 .
\EEQ
For the short-range case, this formula has
already been found in \cite{Calabrese04}.
\item
The spin$^2$--energy-density correlation functions are  
\BEQ
\label{s2-en-corr}
\hspace{-1.5cm}
{\cal C}_{2 \epsilon}(t,s;\vec{r}) = \langle
\mathcal{O}_2(t,\vec{r}) \mathcal{O}_\epsilon(s,\vec{0})
\rangle = \frac{-1}{2 g(t; T)} 
\partial_t \Big( g(t; T)  {\cal C}_{22}(t,s;\vec{r}) \Big) ,
\EEQ
and 
\BEQ
\label{en-s2-corr}
{\cal C}_{\epsilon 2}(t,s;\vec{r}) = {\cal C}_{2 \epsilon}(t,s;\vec{r}).
\EEQ
This is a stronger result than the obvious relation 
${\cal C}_{\epsilon 2}(t,s;\vec{r}) ={\cal C}_{2\epsilon}(s,t;-\vec{r})$
and follows from $\omega(\vec{k})=\omega(-\vec{k})$. 
\item
The energy-density--energy-density correlation function is given by 
\BEQ
\label{en-en-corr}
{\cal C}_{\epsilon \epsilon}(t,s;\vec{r}) = 
\frac{-1}{2 g(t; T)} 
\partial_t
\Big( g(t; T)  {\cal C}_{2 \epsilon}(t,s; \vec{r}) \Big) .
\EEQ
\end{itemize}

\subsubsection{The response functions:} ~\\
For the response functions, we obtain the following expressions. Because of causality, in all expressions given below the factor $\Theta(t-s)$ is implied , 
where the step function $\Theta(t-s) =1$ for $t>s$, and zero otherwise. 
\begin{itemize}
\item
Responses to the magnetic field $h_1(t,\vec{x})$, which are obtained when
${\cal H} \rightarrow {\cal H} - \sum_{t,\vec{x}} h_1(t,\vec{x}) S(t,\vec{x})$, 
are given by
\BEQ
\label{11-resp}
{\cal R}_{11}(t,s;\vec{r}) = 
\sqrt{\frac{g(s; T)}{g(t; T)}} ~f\left(
\frac{t-s}{2},\vec{r}\right) ,
\EEQ
\BEQ
\label{21-en1-resp}
{\cal R}_{21}(t,s;\vec{r}) = 
{\cal R}_{\epsilon1}(t,s;\vec{r}) = 0.
\EEQ
\item
Responses to the conjugate field $h_2(t,\vec{x})$  of spin$^2$ operator are 
obtained when 
${\cal H} \rightarrow {\cal H} - \sum_{t,\vec{x}} h_2(t,\vec{x}) 
{\cal O}_2(t,\vec{x})$ and are given by
\BEQ
\label{12-resp}
{\cal R}_{12}(t,s;\vec{r}) = 0,
\EEQ
\BEQ
\label{22-resp}
{\cal R}_{22}(t,s;\vec{r}) = 
4 {\cal R}_{11}(t,s;\vec{r}) C(t,s;\vec{r}) ,
\EEQ
\BEQ
\label{en2-resp}
{\cal R}_{\epsilon 2}(t,s;\vec{r}) = 
- {\cal R}_{22}(t,s;\vec{r}) 
~\partial_t \ln f\left(\frac{t-s}{2},\vec{r}\right) ,
\EEQ
The expression for ${\cal R}_{22}(t,s; \vec{r})$ has already
been given in \cite{Calabrese04} for the short-range model.
\item
Responses to the conjugate field $h_{\epsilon}(t,\vec{x})$  of 
energy-density operator are obtained when 
${\cal H} \rightarrow {\cal H} - \sum_{t,\vec{x}} 
h_{\epsilon}(t,\vec{x}) {\cal O}_{\epsilon}(t,\vec{x})$ 
and are given by
\BEQ
\label{1en-resp}
{\cal R}_{1\epsilon}(t,s;\vec{r}) = 0,
\EEQ
\BEQ
\label{2en-resp}
{\cal R}_{2 \epsilon}(t,s;\vec{r}) = 
\frac{-1}{2 g(t; T)} 
\partial_t
\Big( g(t; T)  {\cal R}_{22}(t,s;\vec{r}) \Big) ,
\EEQ
\BEQ
\label{enen-resp}
{\cal R}_{\epsilon \epsilon}(t,s;\vec{r}) = 
\frac{-1}{2 g(t; T)} \partial_t \Big( 
g(t; T) {\cal R}_{2 \epsilon}(t,s;\vec{r}) \Big) .
\EEQ
\item
The spin, the spin$^2$ and the energy-density responses to temperature 
fluctuation are
\BEQ
\label{1T-resp}
{\cal R}^{(T)}_{1}(t,s;\vec{r}) = 0,
\EEQ 
\BEQ
\label{2T-resp}
{\cal R}^{(T)}_{2}(t,s;\vec{r}) =
2 \Big( {\cal R}_{11}(t,s;\vec{r}) \Big)^2,
\EEQ 
\BEQ
\label{enT-resp}
{\cal R}^{(T)}_{\epsilon}(t,s;\vec{r}) =
\frac{-1}{2 g(t; T)} 
\partial_t \Big( g(t; T) {\cal R}_{2}(t,s;\vec{r}) \Big),
\EEQ 
respectively. 
\end{itemize}

\subsection{Late-time behaviour of correlation- and response- functions}
\label{largescale}

In this section, we first explicitly evaluate in the scaling limit
the quantities 
specified in the previous subsection, and then identify 
the critical exponents and 
scaling functions. The treatment is based on previous results and 
techniques from \cite{Cann01,Godreche00}. 

In the late-time limit we can approximate the function 
$\omega(\vec{k}) \approx B|\vec{k}|^{\sigma}$, 
where $0 < \sigma < 2$\cite{Cann01}. Hence the dynamical exponent 
in this range of $\sigma$ is given by
\BEQ
z = \sigma .
\EEQ
Furthermore, the large-time behaviour of $f(t)$ and $g(t;T)$ are as follows.  
The function $f(t,\vec{x})$ in this limit becomes
\begin{eqnarray}
\label{f-fn-asym}
f(t;\vec{x}) \approx  
B_0 t^{-d/\sigma} 
G(|\vec{x}|t^{-1/\sigma})\;\; ; \;\; 
B_0 := \int_{\vec{k}} e^{-2 B|\vec{k}|^{\sigma} }.
\end{eqnarray}
Here the scaling function $G(|\vec{\mathfrak{u}}|t^{-1/\sigma})$ for any variable 
$ \vec{\mathfrak{u}}$ is defined as
\BEQ
G(|\vec{\mathfrak{u}}|t^{-1/\sigma}) := 
B_0^{-1} t^{d/\sigma}
\int_{{\vec{k}}} 
e^{\II \vec{k} \cdot \vec{\mathfrak{u}}  } e^{-2 B|\vec{k}|^{\sigma} t} ,
\EEQ
where $\int_{\vec{k}} \cdots = (2\pi)^{-d} \int\!\D^d k \cdots$ denotes an 
integral over $\mathbb{R}^d$.

The Laplace transform of $f(t)$ is given by the expression
\BEQ
\label{f:L}
f_L(p) = - A_0 p^{-1+d/\sigma}  + \sum_{n=1}^{\infty}A_n (-p)^{n-1} ,
\EEQ
where the universal constant
$A_0 = |\Gamma(1-d/\sigma)| B_0$ and
the nonuniversal constants 
$A_n = \int_{\Lambda_{\vec{k}}} (2\omega_{\vec{k}})^{-n} 
- \int_{\vec{k}} (2B |\vec{k}|^{\sigma})^{-n}$, 
for $n=1,2\dots$. We note that $A_1=1/2T_c$. 

Now the constraint equation (\ref{volt}), upon Laplace transforming, becomes 
\BEQ
\label{volt:L}
g_L(p;T) = \frac{f_L(p)} { 1 - 2T f_L(p)},
\EEQ
and is solved in the small-$p$ region using equation (\ref{f:L}). Following 
a similar analysis as done for $\sigma=2$ case in \cite{Godreche00}, we find
the large-$t$ limit of the function $g(t;T)$, which is given in equations 
(\ref{g-fn:I}), (\ref{g-fn:IIa}), and (\ref{g-fn:IIb}).
This asymptotic constraint function has three different forms depending 
on the quenched temperature and the lattice dimension, for a given value of
the parameter $\sigma$. The known case for the short-range
model one can obtain by taking the limit $\sigma \rightarrow
2$. The three cases are
\begin{itemize}
   \item \underline{$T < T_c$:} This case was
   treated in \cite{Cann01} for the spin--spin correlator
   and the spin response. We recover their results and further add
   other correlation and response functions of the composite fields.
   \item \underline{$T = T_c, \sigma < d < 2 \sigma$:} To the best 
   of our knowledge the quench to criticality has not been treated before.
   We must further distinguish two critical cases. In the first case, $d$ can 
   at most be 4 since $\sigma \le 2$. 
   \item \underline{$T = T_c, d > 2 \sigma$:} In this second case of a critical
   quench, the space dimension $d$ is not bounded from
   above. This case includes the mean-field case.
   \end{itemize}
We now discuss the large-time behaviour of the correlation and response 
functions in these three cases.

\subsubsection{\underline{Case I: $T < T_c$}} 
\label{caseI}  ~\\
Since the system exhibits space-translation invariance we take 
$\vec{x}'=\vec{0}$. We denote $y = t/s > 1$. The constraint function for $T < T_c$ in the 
large-time limit \cite{Cann01} is
\BEQ
\label{g-fn:I}
g(t;T) \approx 
B_0
\left( 1 - \frac{T}{T_c} \right)^{-1} t^{-d/\sigma} ,
\EEQ
and hence the spin-spin correlation function for $T < T_c$ in the 
scaling regime reduces to 
\BEQ
\label{C-fn:I}
\wht{C}(t,s;\vec{k}) =  
\left( 1-\frac{T}{T_c} \right)
B_0^{-1}
~s^{d/\sigma} ~ y^{d/2\sigma} e^{-B|\vec{k}|^{\sigma}(t+s)} ,
\EEQ
in the Fourier space, or  
\BEQ
\label{C-fnR:I}
C(t,s;\vec{r}) =  C_0 ~ y^{d/2\sigma} (y+1)^{-d/\sigma} G(u),
\EEQ
in the direct space,
where $ C_0 = 2^{d/\sigma}(1-T/T_c)$. Here and below, expressions become 
shorter with the use of the three related scaling variables $u$, $v$, and 
$w$, where 
\BEA
\label{vdef}
u &=&|\vec{r}|((t+s)/2)^{-1/\sigma} \:=\: w (1+s/t)^{-1/\sigma} \;\; ,  
\nonumber \\
v &=&|\vec{r}|((t-s)/2)^{-1/\sigma} \:=\: w (1-s/t)^{-1/\sigma} \;\; ,
\nonumber \\
w &=& |\vec{r}|(t/2)^{-1/\sigma} \;\; .
\EEA
The autocorrelation function can now be directly deduced since 
the scaling function $G(0) = 1$ for $\vec{r} = 0$. Hence one reads off, 
see (\ref{expC}) and table~\ref{table1}, 
\BEQ
b_{11} = 0, \quad \lambda_{C}^{11} = \frac{d}{2 }, \quad
f_{C}^{11}(y) = C_0 ~ y^{d/2\sigma} (y+1)^{-d/\sigma}.
\EEQ

\begin{table}[t]
\[
\hspace{-2.0cm}
\begin{array}{||c||c|c||c|c|c||} \hline \hline 
 & \multicolumn{2}{|c||}{b} 
 & \multicolumn{3}{|c||}{\lambda_C} \\ \cline{2-6}
 \mbox{Function} & ~~T < T_c~~ & ~~T=T_c~~ 
 & ~~T < T_c~~ & \multicolumn{2}{|c||}{T=T_c} \\
 \cline{5-6}
  &  &  &  & \sigma < d < 2\sigma &  ~~d > 2\sigma~~ \\
  \hline \hline
  {\cal C}_{11} & 0   & d/\sigma-1 & d/2 & 3d/2-\sigma & d\\ 
  \hline 
  {\cal C}_{22} & 0  & 2d/\sigma-2 & d & 3d-2\sigma & 2d\\ 
  \hline 
  {\cal C}_{2\epsilon} & 1   & 2d/\sigma-1 & d+\sigma &
  3d-\sigma & 
  2d+\sigma \\ 
  \hline 
  {\cal C}_{\epsilon\epsilon} & 2  & 2d/\sigma & d+2\sigma &
  3d & 
  2d+2\sigma\\ 
  \hline
  \hline \hline
  \end{array}
  \]
  \caption{\label{table1} Non-equilibrium exponents $b$, $\lambda_C$, 
as defined in (\ref{expC}),
for several non-equilibrium autocorrelation functions in the long-range
spherical model. The exponents for the short-range model can
be recovered by taking the limit $\sigma \rightarrow 2$.
  } 
  \end{table}

\begin{table}[t]
\[
\hspace{-2.0cm}
\begin{array}{||c||c|c||c|c|c||} \hline \hline 
 & \multicolumn{2}{|c||}{a} 
 & \multicolumn{3}{|c||}{\lambda_R} \\ \cline{2-6}
 \mbox{Function} & ~~T < T_c~~ & ~~T=T_c~~ 
 & ~~T < T_c~~ & \multicolumn{2}{|c||}{T=T_c} \\
 \cline{5-6}
  &  &  &  & \sigma < d < 2\sigma &  ~~d > 2\sigma~~ \\
  \hline \hline
  \hline
  {\cal R}_{11} &d/\sigma-1   & d/\sigma-1  & d/2  &
  3d/2-\sigma &
  d \\ 
  \hline 
  {\cal R}_{22} & d/\sigma-1  & 2d/\sigma-2  & d &
  3d-2\sigma & 
  2d \\ 
  \hline 
  {\cal R}_{\epsilon 2} & d/\sigma  & 2d/\sigma-1 & d
  +\sigma & 
  3d-\sigma & 2d+\sigma\\ 
  \hline 
  {\cal R}_{2\epsilon} & d/\sigma  & 2d/\sigma-1 & d+\sigma
  & 
  3d-\sigma & 2d+\sigma\\ 
  \hline 
  {\cal R}_{\epsilon\epsilon} & d/\sigma+1  & 2d/\sigma &
  d+2\sigma & 
  3d & 2d+2\sigma\\ 
  \hline 
  {\cal R}^T_{2} & 2d/\sigma-1  & 2d/\sigma-1 & d & 
  3d-2\sigma & 2d\\ 
  \hline 
  {\cal R}^T_{\epsilon} & 2d/\sigma  & 2d/\sigma & d+\sigma
  & 
  3d-\sigma & 2d+\sigma\\ 
  \hline \hline
  \end{array}
  \]
  \caption{\label{table2} Nonequilibrium exponents $a=a'$ and $\lambda_R$,
as defined in (\ref{expR}) and (\ref{expRT}), 
for several scaling operators in the long-range spherical
model. The exponents for the short-range model can be obtained
by taking the limit $\sigma \rightarrow 2$.
  } 
  \end{table}

Below we list the remaining expressions in the scaling limit.
The autocorrelation and autoresponse functions are 
obtained for the composite operators in a similar way as is demonstrated for
$C(t,s;\vec{r})= {\cal C}_{11}(t,s;\vec{r})$. The
non-equilibrium ageing exponents are listed in 
tables~\ref{table1} and~\ref{table2}, for future reference.

We first list the non-vanishing correlation functions.
\begin{itemize}
\item The spin$^2$ -- spin$^2$ correlator, obtained by
substituting equation (\ref{C-fnR:I})
into (\ref{s2-s2-corr}), is
\BEQ
\label{s2-s2-corr:I}
{\cal C}_{22}(t,s;\vec{r}) = 
2 C_0^2 ~y^{d/\sigma} (y+1)^{-2d/\sigma} G^2(u).
\EEQ
\item The spin$^2$ -- energy-density correlator, obtained
by using equations (\ref{g-fn:I}, \ref{s2-s2-corr:I})
in (\ref{s2-en-corr}), is
\BEQ
\label{s2-en-corr:I}
{\cal C}_{2 \epsilon}(t,s;\vec{r}) = \frac{2C_0^2} {\sigma}
~ s^{-1} y^{d/\sigma} (y+1)^{-1-2d/\sigma}  G(u)D_uG(u),
\EEQ
where, the operator $D_z$ is defined as
\BEQ
\label{deriv}
D_z := z \partial_z + d .
\EEQ
\item The energy-density -- energy-density correlator,
obtained by inserting equations (\ref{g-fn:I}, \ref{s2-en-corr:I})
into (\ref {en-en-corr}), is given by
\BEQ
\hspace{-1.5cm}
\label{en-en-corr:I}
{\cal C}_{\epsilon \epsilon}(t,s;\vec{r}) = \frac{C_0^2} {\sigma^2}
~ s^{-2} y^{d/\sigma} (y+1)^{-2-2d/\sigma} 
\left( D_u + d + \sigma\right)[G(u)D_uG(u)].
\EEQ
\end{itemize}
Next we write down the non-vanishing response functions.
\begin{itemize}
\item
The spin response function, obtained using equations (\ref{f-fn-asym}, \ref{g-fn:I})
in (\ref{11-resp}), is given by
\BEQ
\label{11-resp:I}
{\cal R}_{11}(t,s;\vec{r}) = C_1 ~
s^{-d/\sigma} y^{d/2\sigma} (y-1)^{-d/\sigma} G(v),
\EEQ
where $C_1 = \int_{\vec{k}}\exp(-B|\vec{k}|^{\sigma})$, and $v$ was defined in eq.~(\ref{vdef}).
\item The non-vanishing response functions to spin$^2$ conjugate field, inferred 
from equations (\ref{22-resp}, \ref{en2-resp}) using 
(\ref{f-fn-asym}, \ref{C-fnR:I}, \ref{11-resp:I}),  
are given by
\BEQ
\label{22-resp:I}
{\cal R}_{22}(t,s;\vec{r}) = 4 C_0 C_1 
~s^{-d/\sigma} y^{d/\sigma} (y^2-1)^{-d/\sigma}  G(u)G(v),
\EEQ
\BEQ
\label{en2-resp:I}
{\cal R}_{\epsilon 2}(t,s;\vec{r}) = \frac{4 C_0 C_1}{\sigma} 
~s^{-1-d/\sigma} y^{d/\sigma} (y^2-1)^{-d/\sigma}  
\frac{D_v}{y-1} G(u)G(v),
\EEQ
where $D_v$ is as given in (\ref{deriv}) and $u,v$ were
defined in (\ref{vdef}).
\item Responses to the energy-density conjugate field, obtained from
equations (\ref{2en-resp}, \ref{enen-resp}) using (\ref{g-fn:I}, \ref{22-resp:I}),
are given as follows.
\begin{eqnarray}
\label{2en-resp:I}
{\cal R}_{2 \epsilon}(t,s;\vec{r}) &=& \frac{2 C_0 C_1}{\sigma}
~ s^{-1-d/\sigma} y^{d/\sigma} (y^2-1)^{-d/\sigma}  \nonumber \\
& & \times \left( \frac{D_u}{y+1}+\frac{D_v}{y-1} \right)G(u)G(v),
\end{eqnarray}
\begin{eqnarray}
\label{enen-resp:I}
&& \hspace{-1.5cm} {\cal R}_{\epsilon \epsilon}(t,s;\vec{r}) 
= \frac{C_0 C_1}{\sigma^2}
~ s^{-2-d/\sigma} y^{d/\sigma} (y^2-1)^{-d/\sigma} \nonumber \\
& & \hspace{-0.5cm} \times \left(  \frac{D_u^2 + \sigma D_u}{(y+1)^2} 
+ \frac{2D_u D_v}{y^2-1} 
+ \frac{D_v^2 + \sigma D_v}{(y-1)^2} \right) G(u)G(v).
\end{eqnarray}
\item The spin$^2$ and energy-density responses to local temperature 
fluctuations, obtained using equations (\ref{g-fn:I}, \ref{11-resp:I})
in (\ref{2T-resp}, \ref{enT-resp}), are
\BEQ
\label{2T-resp:I}
{\cal R}_{2}^{(T)}(t,s;\vec{r}) = 2 C_1^2
~s^{-2d/\sigma} y^{d/\sigma} (y-1)^{-2d/\sigma} G^2(v),
\EEQ 
\BEQ
\label{enT-resp:I}
{\cal R}_{\epsilon}^{(T)}(t,s;\vec{r}) = \frac{2C_1^2}{\sigma}
~s^{-1-2d/\sigma} y^{d/\sigma} (y-1)^{-1-2d/\sigma} G(v)D_vG(v),
\EEQ 
respectively.
\end{itemize}

\subsubsection{\underline{Case IIa: $T = T_c$ and $\sigma < d < 2\sigma$}}
\label{caseIIa} ~\\
For $T=T_c$ and  $\sigma < d < 2\sigma$, the constraint function has the form
\BEQ
\label{g-fn:IIa}
g(t;T_c) \approx \left( 4T_c^2 A_0 \Gamma(-1+d/\sigma)\right)^{-1} 
t^{-2 + d/\sigma} ,
\EEQ
and hence the correlation function in the scaling regime reduces to 
\BEQ
\label{C-fn:IIa}
\wht{C}(t,s;\vec{k}) =  2 T_c~s~y^{1-d/2\sigma}
\int_0^1\!dz~ e^{- B|\vec{k}|^{\sigma}(t+s-2sz)}  z^{-2+d/\sigma} ,
\EEQ
while in direct space is given by
\BEQ
\label{C-fnR:IIa}
C(t,s;\vec{r}) =  2 T_c C_1~s^{1-d/\sigma}~y^{1-d/2\sigma} (y+1)^{-d/\sigma}
\sum_{n=0}^{\infty} \frac{(y+1)^{-n} G_n(u)}{n!(n-1+d/\sigma)} ,
\EEQ
where $u$ is given in (\ref{vdef}) and the function
$G_n(|\vec{\mathfrak{v}}|t^{-1/\sigma})$ is defined as 
\BEQ
\label{G_n:def}
G_n(|\vec{\mathfrak{v}}|t^{-1/\sigma}) :=  
4^n t^{n+d/\sigma} B_0^{-1}
\int_{{\vec{k}}} 
e^{\II \vec{k} \cdot \vec{\mathfrak{v}} } e^{-2B|\vec{k}|^{\sigma}t} 
( B|\vec{k}|^{\sigma})^n, 
\EEQ
for any variable $\vec{\mathfrak{v}}$.
The spin-response function in this case has the form 
\BEQ
\label{spin-resp:IIa}
{\cal R}_{11}( t, s;\vec{x}) = C_1 ~s^{-d/\sigma} y^{1-d/2\sigma} 
(y-1)^{-d/\sigma} G(v).
\EEQ
To avoid presenting lengthy expressions we write down only the leading 
behaviour in $y$ for the correlators and responses in this case. The 
spin-spin correlation 
function in this approximation becomes
\BEQ
\label{C-fnR-approx:IIa}
C(t,s;\vec{r}) \approx 2\widetilde{T}_c C_1  
~s^{1-d/\sigma} y^{1-3d/2\sigma} G(w), 
\EEQ
where $\widetilde{T}_c= T_c\sigma/(d-\sigma)$, and $w$ is as given in (\ref{vdef}). 
Setting $w$ and $v$ to zero, we can read off the ageing
exponents, see tables~\ref{table1} and~\ref{table2}.
\BEQ
a_{11} = b_{11} = \frac{d}{\sigma} -1, \qquad  \lambda_{R}^{11} =
\lambda_{C}^{11} = \frac{3 d}{2}- \sigma, \qquad z = \sigma
\EEQ

The other non-vanishing correlators and responses are given as 
follows, wherein we first list the correlation functions.
\begin{itemize}
\item The spin$^2$ -- spin$^2$ correlator, obtained from equations
(\ref{C-fnR-approx:IIa}, \ref{s2-s2-corr}), is given by
\BEQ
\label{s2-s2-corr:IIa}
{\cal C}_{22}(t, s; \vec{r}) \approx 8\widetilde{T}_c^2 C_1^2 
~s^{2-2d/\sigma} y^{2-3d/\sigma} G^2(w).
\EEQ
\item For the spin$^2$ -- energy correlator,
using (\ref{g-fn:IIa}, \ref{s2-s2-corr:IIa}) in (\ref{s2-en-corr}), we obtain
\BEQ
\label{s2-en-corr:IIa}
{\cal C}_{2 \epsilon}(t,s;\vec{r}) \approx 
\frac{8\widetilde{T}_c^2 C_1^2}{\sigma} 
~s^{1-2d/\sigma} y^{1-3d/\sigma} G(w)D_wG(w).
\EEQ
\item Finally the energy-- energy correlator, using
(\ref{g-fn:IIa}, \ref{s2-en-corr:IIa}) in  (\ref{en-en-corr}), reads
\BEQ
\label{en-en-corr:IIa}
{\cal C}_{\epsilon \epsilon}(t, s; \vec{r}) \approx 
\frac{4\widetilde{T}_c^2 C_1^2}{\sigma^2}
~s^{-2d/\sigma} y^{-3d/\sigma} (D_w+d+\sigma)[G(w)D_wG(w)].
\EEQ
\end{itemize}
The non-vanishing response functions are listed below.
\begin{itemize}
\item The responses to the spin$^2$ conjugate field, obtained using
(\ref{f-fn-asym}, \ref{spin-resp:IIa}, \ref{C-fnR-approx:IIa})
in (\ref{22-resp}, \ref{en2-resp}), are given by
\BEQ
\label{22-resp:IIa}
{\cal R}_{22}(t, s; \vec{r}) \approx 8\widetilde{T}_c C_1^2 
~s^{1-2d/\sigma} y^{2-3d/\sigma} G^2(w),
\EEQ
\BEQ
\label{en2-resp:IIa}
{\cal R}_{\epsilon 2}( t, s ; \vec{r}) \approx 
\frac{8\widetilde{T}_c C_1^2}{\sigma} ~s^{-2d/\sigma} y^{1-3d/\sigma} 
G(w)D_wG(w).
\EEQ
\item The responses to energy-density conjugate field, obtained from
(\ref{g-fn:IIa}, \ref{22-resp:IIa}) and (\ref{2en-resp}, \ref{enen-resp}), are
\BEQ
\label{2en-resp:IIa}
{\cal R}_{2 \epsilon}(t, s; \vec{r}) \approx {\cal R}_{\epsilon 2}( t, s ; \vec{r}),  
\EEQ
\BEQ
\label{enen-resp:IIa}
{\cal R}_{\epsilon \epsilon}(t, s; \vec{r}) \approx 
\frac{4\widetilde{T}_c C_1^2}{\sigma^2} 
~s^{-1-2d/\sigma} y^{-3d/\sigma} (D_w +d +\sigma )[G(w)D_wG(w)].
\EEQ
\item Lastly, the responses to temperature fluctuations, obtained from
(\ref{g-fn:IIa}, \ref{spin-resp:IIa}) and (\ref{2T-resp}, \ref{enT-resp}), are
\BEQ
\label{2T-resp:IIa}
{\cal R}_{2}^{(T)}(t, s; \vec{r}) \approx 
2 C_1^2 ~s^{-2d/\sigma} y^{2-3d/\sigma} G^2(w),
\EEQ 
\BEQ
\label{enT-resp:IIa}
{\cal R}_{\epsilon}^{(T)}(t, s; \vec{r}) \approx 
\frac{2 C_1^2}{\sigma} ~s^{-1-2d/\sigma} y^{1-3d/\sigma} G(w)D_wG(w).
\EEQ 
\end{itemize}

\subsubsection{\underline{Case IIb: $T = T_c$ and $d > 2\sigma$}}
\label{caseIIb} ~\\
For $T=T_c$ and  $d > 2\sigma$, the constraint function at large times is
\BEQ
\label{g-fn:IIb}
g(t;T_c) \approx \left( 4T_c^2 A_2  \right)^{-1} .
\EEQ
This is just a constant and does not appear in the correlation and response functions to 
leading order in this large-time limit. In this case, the correlation function in the 
scaling regime reduces to 
\BEQ
\label{C-fn:IIb}
\wht{C}(t,s;\vec{k}) =  \frac{T_c}{B|\vec{k}|^{\sigma}}
\left( e^{-B|\vec{k}|^{\sigma}(t-s)}  - e^{-B|\vec{k}|^{\sigma}(t+s)} \right) ,
\EEQ
and in the direct space is
\BEQ
\label{C-fnR:IIb}
C(t, s; \vec{r}) =  2T_c C_1 ~ s^{1-d/\sigma}
\left(
\frac{G_{-1}(v)}{(y-1)^{d/\sigma -1}}  - \frac{G_{-1}(u)}{(y+1)^{d/\sigma-1}}
\right),
\EEQ
where $G_{-1}$ is as given in (\ref{G_n:def}).

The spin-response function in this case is given by
\BEQ
\label{spin-resp:IIb}
{\cal R}_{11}(t, s; \vec{r}) = C_1 ~s^{-d/\sigma} (y-1)^{-d/\sigma} G(v).
\EEQ
Here again we present only the leading behaviour in $y$ of the correlators and 
responses. The correlation function in this approximation becomes
\BEQ
\label{C-fnR-approx:IIb}
C(t, s; \vec{r}) \approx  2T_c~ s~ f(t/2, \vec{r})= 2T_c C_1 
~s^{1-d/\sigma} y^{-d/\sigma} G(w).
\EEQ
Again we read off the critical exponents after setting $v = w = 0$
\BEQ
a_{11} = b_{11} = \frac{d}{\sigma}-1, \quad \lambda_{R}^{11} =
\lambda_{C}^{11} = d .
\EEQ
The other non-vanishing correlation functions are given as follows.
\begin{itemize}
\item The spin$^2$--spin$^2$ correlation function, substituting
(\ref{C-fnR-approx:IIb}) in (\ref{s2-s2-corr}), is
\BEQ
\label{s2-s2-corr:IIb}
{\cal C}_{22}(t, s; \vec{r}) \approx 8T_c^2 C_1^2 
~s^{2-2d/\sigma} y^{-2d/\sigma} G^2(w).
\EEQ
\item The spin$^2$ -energy correlation function,
from (\ref{g-fn:IIb}, \ref{s2-s2-corr:IIb}, \ref{s2-en-corr}),
\BEQ
\label{s2-en-corr:IIb}
{\cal C}_{2 \epsilon}(t, s; \vec{r}) \approx 
\frac{8T_c^2 C_1^2}{\sigma} ~s^{1-2d/\sigma} y^{-1-2d/\sigma} G(w)D_wG(w).
\EEQ
\item The energy-density -- energy-density correlation function,
from (\ref{g-fn:IIb}, \ref{s2-en-corr:IIb}, \ref{en-en-corr}), is
\BEQ
\label{en-en-corr:IIb}
{\cal C}_{\epsilon \epsilon}(t, s; \vec{r}) \approx 
\frac{4T_c^2 C_1^2}{\sigma^2} 
~s^{-2d/\sigma} y^{-2-2d/\sigma} (D_w+d+\sigma)[G(w)D_wG(w)] .
\EEQ
\end{itemize}
The remaining non-vanishing response functions follow.
\begin{itemize}
\item The responses to spin$^2$ conjugate field, obtained from
(\ref{f-fn-asym}, \ref{spin-resp:IIb}, \ref{C-fnR-approx:IIb})
and (\ref{22-resp}, \ref{en2-resp}), are
\BEQ
\label{22-resp:IIb}
{\cal R}_{22}(t, s; \vec{r}) \approx
8T_c C_1^2 ~s^{1-2d/\sigma} y^{-2d/\sigma} G^2(w),
\EEQ
\BEQ
\label{en2-resp:IIb}
{\cal R}_{\epsilon 2}(t, s; \vec{r}) \approx \frac{8T_c C_1^2}{\sigma} 
~s^{-2d/\sigma} y^{-1-2d/\sigma} G(w)D_wG(w).
\EEQ
\item The responses to energy-density conjugate field, otained from
(\ref{g-fn:IIb}, \ref{22-resp:IIb}) and (\ref{2en-resp}, \ref{enen-resp}),
are given by
\BEQ
\label{2en-resp:IIb}
{\cal R}_{2 \epsilon}(t, s; \vec{r}) \approx 
{\cal R}_{\epsilon 2}(t, s; \vec{r}) ,
\EEQ
\BEQ
\label{enen-resp:IIb}
{\cal R}_{\epsilon \epsilon}( t, s; \vec{r}) \approx 
\frac{4T_c C_1^2}{\sigma^2} 
~s^{-1-2d/\sigma} y^{-2-2d/\sigma} (D_w +d +\sigma )[G(w)D_wG(w)] .
\EEQ
\item Finally, the responses to temperature fluctuations, obtained from
(\ref{g-fn:IIb}, \ref{spin-resp:IIb}) and (\ref{2T-resp}, \ref{enT-resp}),
are given as
\BEQ
\label{2T-resp:IIb}
{\cal R}_{2}{(T)}(t, s; \vec{r}) \approx
2 C_1^2 ~s^{-2d/\sigma} y^{-2d/\sigma} G^2(w),
\EEQ 
\BEQ
\label{enT-resp:IIb}
{\cal R}_{\epsilon}^{(T)}(t,s; \vec{r}) \approx \frac{2 C_1^2}{\sigma} 
~s^{-1-2d/\sigma} y^{-1-2d/\sigma} G(w)D_wG(w).
\EEQ 
\end{itemize}
The exponents of these functions, derived in this
section, are collected in tables~\ref{table1} and~\ref{table2}.

\subsubsection{\underline{Fluctuation-dissipation ratios}}~\\
An important quantity, in particular for the case of 
critical dynamics, is the fluctuation-dissipation ratio of
an observable, which is defined as \cite{Cugliandolo94,Crisanti03}
\BEQ
X_{ab}(t,s) :=  T_c {\cal
R}_{ab}(t,s;\vec{0})\left(\frac{\partial C_{ab}(t,s;
\vec{0})}{\partial s} \right)^{-1}
\EEQ
and its limit value
\BEQ
X_{ab}^\infty := \lim_{s \rightarrow \infty} \left( \lim_{t
\rightarrow \infty}X_{ab}(t,s) \right) =\lim_{y\to\infty}
\left( \lim_{s\to\infty} \left. X_{ab}(t,s)\right|_{y=t/s} \right).
\EEQ

For case $I$, that is for phase-ordering kinetics, it was already known
that in the quasi-static limit $s\to\infty$ but $t-s$ fixed and $\ll s$, the 
fluctuation-dissipation theorem still holds \cite{Cann01}. On the other
hand, we obtain in the scaling limit $s\to\infty$ and $y=t/s>1$ fixed that,
for all observables considered here
\BEQ
\hspace{-1.5cm}
X_{11}(t,s) = X_{22}(t,s) = X_{2\epsilon}(t,s) =
X_{\epsilon 2}(t,s) = X_{\epsilon \epsilon}(t,s) = \frac{2
\sigma T C_1}{d C_0} s^{1-d/\sigma}.
\EEQ
For $d > \sigma$ we have therefore in this case that
\begin{eqnarray}
X_{11}^\infty = X_{22}^\infty = X_{2\epsilon}^\infty =
X_{\epsilon 2}^\infty = X_{\epsilon \epsilon}^\infty = 0
\end{eqnarray}
as expected for a low-temperature phase 
(recall that for $d \leq \sigma$  the critical temperature is
zero \cite{Joyce66}).

In the case of critical dynamics (case $IIa$ and $IIb$) the
limit fluctuation-dissipation ratios are universal 
numbers characterising the critical system \cite{Godreche00}. For
their calculation, we can use directly the scaling limit
$s\to\infty$ with $y=t/s$ being kept fixed. 
In case IIa, it is convenient to obtain the auto-correlation function $C(t,s)$ 
by directly integrating eq.~(\ref{C-fn:IIa}),  which leads to 
\begin{equation}
\label{autoC}
C(t,s) = \frac{2 T_c C_1\sigma}{d-\sigma}~s^{1-d/\sigma}~y^{1-d/2\sigma} 
(y-1)^{1-d/\sigma} (y+1)^{-1}.
\end{equation}
Combining this with eq.~(\ref{spin-resp:IIa}), we get
\begin{eqnarray}
X_{11}(t,s) = X_{11}(y)= \frac{1}{2} (y+1) 
\left[ 1+ \frac{y-1}{d-\sigma} \left( \frac{d}{2}-\frac{\sigma}{y+1} \right) 
\right]^{-1}
\end{eqnarray}
Similarly, in case IIb, using equations (\ref{C-fnR:IIb}) and 
(\ref{spin-resp:IIb}), and upon substituting the 
value of $G_{-1}(0)= \sigma G(0)/(d-\sigma)$, we find
\begin{eqnarray}
X_{11}(t,s) = X_{11}(y) = \left( 1 + \left(\frac{y-1}{y+1}\right)^{d/\sigma} \right)^{-1}
\end{eqnarray}
In particular, we see that in the quasi-static limit $s\to\infty$ 
with $t-s$ being kept fixed (or alternatively $y\to 1$), 
$\lim_{y\to 1}X_{11}(y) \rightarrow 1$ in both critical cases, such that the 
fluctuation-dissipation theorem holds. 
Similarly, from the relations
(\ref{s2-s2-corr},\ref{s2-en-corr},\ref{en-en-corr}) and
(\ref{22-resp},\ref{2en-resp},\ref{enen-resp}) 
we also have $\lim_{y\to 1} X_{22}(y) = \lim_{y\to 1} 
X_{\epsilon \epsilon}(y) = \lim_{y\to 1} X_{2\epsilon}(y) =1$.
On the other hand, and remarkably, the limit
fluctuation-dissipation ratio turns out to be independent of
the choice of the considered observable. We find for $y\to\infty$ 
\BEQ
X_{11}^\infty = X_{22}^\infty = X_{2\epsilon}^\infty =
X_{\epsilon 2}^\infty = X_{\epsilon \epsilon}^\infty =\left\{ \begin{array}{cc}
1 -{\sigma}/{d} & \mbox{for the case $IIa$} \\
{1}/{2} & \mbox{for the case $IIb$} \end{array} \right.
\EEQ
This reduces to the well-known expressions in the short-range
model \cite{Godreche00} when $z =\sigma\rightarrow 2$.
We recall that in \cite{Calabrese04}, a slightly different
definition for the energy density was used, in which case
the value for the corresponding fluctuation-dissipation
ratio may be different. 

\section{Local scale-invariance}

The theory of local scale-invariance (LSI) was developed in a
series of papers \cite{Henkel94,Henkel02,Picone04,Roethlein06,Henkel06a}, using
local symmetries to fix the response and correlation functions.
For recent reviews which focus on different types of
applications see \cite{Henkel07a,Henkel07b,Henkel07c}. For our
purposes here it is sufficient to just quote a few results.
A central concept of LSI are the {\it quasi-primary} scaling operators 
\cite{Henkel02}, which transform in the simplest possible way under local 
scale-transformations, very much in analogy with the (quasi)primary scaling 
operators of conformal field-theory \cite{Bela84}.\footnote{Specifically,
if ${\cal X}$ is an infinitesimal generator of a local scale-transformation and
$\phi$ a quasi-primary scaling operator, $\delta \phi = -\eps {\cal X}\phi$. Usually, the order-parameter corresponds to a quasi-primary operator, but if
$\phi$ is quasi-primary, then neither $\partial_t \phi$ nor $\partial_{\vec{r}}\phi$ are. The $n$-point functions 
$\langle \phi_1\ldots\phi_n\rangle$ of quasi-primary operators 
transform covariantly and hence satisfy linear differential equations 
${\cal X}^{[n]}\langle \phi_1\ldots\phi_n\rangle=0$.} A
quasi-primary scaling operator $\phi$ is characterised by a set of `quantum
numbers' $(x,\xi,\mu,\beta)$, where $x$ is the `scaling
dimension' of $\phi$ and $\mu$ is sometimes referred to as the `mass' of 
$\phi$ (not to be confused with the lattice constant $\mu$ in section~2.1). 

\subsection{Response functions}

For a given dynamical exponent $z$, 
LSI yields the following prediction for the response
function of a quasi-primary operator $\phi$ characterised by the 
parameters $(x,\xi,\mu,\beta)$:
\cite{Henkel02,Roethlein06,Baumann07,Henkel07c}
\BEA
R^{LSI}(t,s;\vec{r})  &=& 
\delta_{\mu,-\tilde{\mu}}\,
\delta_{\beta, \tilde{\beta}}\,
R(t,s)\mathcal{F}^{(\mu,\beta)}
\left(\frac{|\vec{r}|}{(t-s)^{1/z}} \right),
\nonumber \\
R(t,s) &=& s^{-1-a}\left(\frac{t}{s} \right)^{1+a'-\lambda_R/z}
\left(\frac{t}{s}-1 \right)^{-1-a'},
\label{response_lsi}
\EEA
where the exponents $a, a'$ and $\lambda_R$ are related to
the parameters $(x,\xi,\mu)$ via
\BEQ
a+1 = \frac{1}{z}(x+\tilde{x}), \quad a' +1 = \frac{1}{z}(x
+ 2 \xi + \tilde{x} + 2 \tilde{\xi}), \quad \frac{\lambda_R}{z}
= \frac{2 x}{z}+ \frac{2 \xi}{z},
\EEQ
and the parameters ($\tilde{x}, \tilde{\xi},\tilde{\mu}, \tilde{\beta}$) 
characterise the response field
$\tilde{\phi}$.
The space-time part $\mathcal{F}^{(\mu,\beta)}(\rho)$ (where
$\rho:=
|\vec{\rho}|$ and $\vec{\rho} = \vec{r}(t-s)^{-1/\sigma}$)
satisfies the following fractional differential equation 
\BEQ \label{eqd_vieille}
\left( \partial_\rho + z \mu \,\rho \partial_\rho^{2-z} + [\beta \mu +
\mu (2 -z)] \partial_\rho^{1-z} \right)
\mathcal{F}^{(\mu,\beta)}(\rho) = 0.
\EEQ
which also illustrates that the `mass' $\mu$ may be interpreted as a generalised diffusion constant. 
The fractional derivatives $\partial_{\rho}^{\alpha}$ 
are defined and discussed in
\cite{Henkel02}. Recall, however, that the definition used here is not
unique and that different non-equivalent definitions for
fractional derivatives exist \cite{Miller93,Podlubny99}.  
If $z=N+p/q$, where $N=[z]$ is the largest integer less or equal to $z$,
$0\leq p/q<1$ and $p$ and $q$ coprime, the solution of (\ref{eqd_vieille})
by series methods is particularly simple, with the result \cite{Roethlein06}
\BEQ
\label{lsi_old}
\hspace{-1.0cm}
\mathcal{F}^{(\mu,\beta)}(\rho) = \sum_{m \in
\mathcal{E}} c_m \phi^{(m)}(\rho), \quad \mbox{with} \quad
\phi^{(m)}(\rho) = \sum_{n = 0}^\infty b_n^{(m)} \rho^{(n-1) z +
p/q + m + 1}.
\EEQ
The
constants $c_m$ are not determined by LSI and the set $\mathcal{E}$ is 
\BEQ
\mathcal{E} = \left\{ \begin{array}{cc}
                      -1,0,\ldots,N-1 &  p \neq 0 \\
		      0,\ldots,N-1, & p = 0
		      \end{array} \right. .
\EEQ
Finally, the coefficients $b_n^{(m)}$ read
\BEQ
b_n^{(m)} = \frac{(-z^2 \mu)^n \Gamma(p/q + 1 +m) \Gamma(n +
z^{-1}(p/q +m) + \beta+2-z)}{\Gamma((n-1) z + p/q + m
+2) \Gamma(z^{-1} (p/q +m ) + \beta + 2 - z)} .
\EEQ
such that $\phi^{(m)}(\rho)$ has an infinite radius of convergence for $z>1$. 

Let us now consider the magnetic response of the order-parameter, $\mathcal{R}_{11}$, the result for which we
recall from (\ref{11-resp}) is
\BEA
\hspace{-1.0cm} \mathcal{R}_{11}(\vec{r};t,s) &=& 
(2 \pi)^d s^{-d/\sigma} \left(\frac{t}{s}\right)^{-\digamma/2}
\left(t/s -1 \right)^{-d/\sigma} \int_{\vec{k}}  
e^{\II \vec{k}\cdot \vec{r}(t-s)^{-1/\sigma}} e^{-B k^\sigma} 
\nonumber \\
\hspace{-1.0truecm}&=& R(t,s) \sum_{n=0}^{\infty} a_n \rho^{2n} 
\;\; , \;\; \vec{\rho}=\vec{r}(t-s)^{-1/\sigma} ,
\label{result_response}
\EEA
where the exponent $\digamma$ is given by
\BEQ \label{exp_alpha}
\digamma = \left\{
\begin{array}{cl} -d/\sigma& \mbox{case (I)} \\
-2 + d/\sigma& \mbox{case (IIa)} \\
0 & \mbox{case (IIb)}
\end{array}
\right..
\EEQ
Clearly, the space-time part of the LSI-prediction does {\em not}
agree with this result since the exponents of $\rho$ in eqs.~(\ref{result_response}) and (\ref{lsi_old}) are linearly independent
if $z$ is not an integer. 
In eq.~(\ref{result_response}), we have expanded 
the exponential in order to rewrite this as a series in
$\rho=|\vec{\rho}|$. 
This form of the series is incompatible with the expected form
(\ref{lsi_old}) for $z<2$. 
This disagreement has motivated us 
to look for a new formulation of LSI, which uses a more
appropriate form of fractional derivatives $\nabla_{r}^\alpha$.
This formulation, including the exact definition of
$\nabla_{r}^\alpha$ will be
described elsewhere in detail \cite{Baumann07}, here we only
mention two results we need:

\noindent 1. {\bf Generalised Bargmann superselection rule:} 
{\it Let a system be given with dynamical exponent $z \neq \frac{2k
+2}{2k+1}$, ($k \in \mathbb{N}$). Let $\{ \phi_i \}$ be a
set of quasi-primary scaling operators, each characterised by the set
$(x_i,\xi_i,\mu_i,\beta_i)$. Then the $(2n)$-point function}
\BEQ
F^{(2 n)} := \langle \phi_1(t_1,\vec{r}_1) \ldots
\phi_{2n}(t_{2n},\vec{r}_{2n}) \rangle.
\EEQ
{\it is zero unless the $\mu_i$ form $n$ distinct pairs
$(\mu_i,\mu_{\tau(i)})$ ($i = 1, \ldots n$), such that}
\BEQ
\mu_i = - \mu_{\tau(i)}.
\EEQ

\noindent 2. The decomposition (\ref{response_lsi}) of the response function 
remains valid, but its space-time part now satisfies the
fractional differential equation, which is quite similar to 
eq.~(\ref{eqd_vieille})
\BEQ \label{eqd_nouvelle}
\left( \partial_\rho + z \mu \, \rho \nabla_{\rho}^{2-z} + [\beta \mu +
\mu (2 -z)] \partial_\rho \nabla_\rho^{-z} \right)
\mathcal{F}^{(\mu,\beta)}(\rho) = 0.
\EEQ

A solution of equation (\ref{eqd_nouvelle}) reads \cite{Baumann07}
\BEQ
\label{spacetime_lsi}
\mathcal{F}^{(\mu,\beta)}(\vec{\rho}) = f_0 \int_{\vec{k}}
e^{\II \vec{\rho} \cdot \vec{k}} \; |\vec{k}|^{\beta}
\exp \left(-\frac{1}{z^2 \II^{2-z} \mu} |\vec{k}|^z
\right)
\EEQ
We see that this prediction of the `new'
formulation of LSI is fully compatible with our exact result
(\ref{result_response}) for
$\mathcal{R}_{11}(t,s;\vec{r})$ if we identify
\BEQ
\label{choice_parameter1}
\mu_1 = -\tilde{\mu}_1 =  (z^2 B \II^{2-z})^{-1}, \quad
\beta_1 = \tilde{\beta}_1 = 0, \quad g_0 = (2 \pi)^d.
\EEQ
and set for the critical exponents
\BEQ
\label{choice_parameter}
a_{11} = a_{11}' = \frac{d}{\sigma}-1, \quad
\lambda^{11}_{R} = d+ \frac{\alpha \sigma}{2}.
\EEQ
This agreement supports the assumption that the fields $\phi$ and
$\wit{\phi}$ are both quasi-primary with $\mu = -
\tilde{\mu}$ and $\beta = \tilde{\beta}$. This is
further supported by the fact that
$\mathcal{R}_{12}(t,t'; \vec{r}) = 0 =
\mathcal{R}_{1 \epsilon }(t,t'; \vec{r})$, which is
predicted by LSI because of the generalised Bargmann
superselection rule.

Having verified that the response function for the order-parameter field
$\phi$ agrees with LSI, and thus having confirmed that $\phi$ is 
indeed quasi-primary, we now inquire whether
this holds for composite operators.  
First, we consider the short-range model $\sigma\geq 2$.
The relevant results can be read from those of section~2 if
we let $\sigma \rightarrow 2$. Then the response
$\mathcal{R}_{11}(t,s;\vec{r})$ in eq.~(\ref{result_response}) simplifies to 
\BEQ
\hspace{-1.0cm}
\mathcal{R}_{11}(t,s;\vec{r}) = 
s^{-d/2} \left( \frac{t}{s}
\right)^{-\digamma/2} \left(\frac{t}{s}-1\right)^{-d/2} 
\exp\left(-\frac{1}{4 B} \frac{\vec{r}^2}{t-s} \right),
\EEQ
up to a normalisation constant.
Similarly, the temperature response of the spin$^2$ field, from the above expression and eq.~(\ref{2T-resp}), becomes
\BEQ
\mathcal{R}_2^{(T)}(t,s;\vec{r}) =  s^{-d}\left(\frac{t}{s} \right)^{-\digamma} 
\left(\frac{t}{s}-1\right)^{-d}
\exp\left(-\frac{1}{2 B} \frac{\vec{r}^2}{t-s} \right),
\EEQ
which is of the form predicted by eq.~(\ref{spacetime_lsi}),
if we identify
\BEQ
\mu_2 = -\tilde{\mu}_2 = 2 \mu_1, \quad \beta_2 =
\tilde{\beta}_2 = 0,
\EEQ
and
\BEQ
a_{22} = a_{22}' = 2 a_{11} + 1, \quad \lambda^{22}_{R} =
2 \lambda^{11}_{R}.
\EEQ
Physically, we can therefore identify temperature changes as the conjugate variable to the spin-squared operator, at least for the short-ranged case. 
On the other hand, the spin$^2$ response
$\mathcal{R}_{22}$ to the
perturbation $h_2(t,\vec{x})$ cannot be cast into that
form. This can easily be seen in equation (\ref{22-resp:I}),
which has a dependence on $t+s$, while the LSI-predicted
form does not contain this dependence.
Note that this response function in a field-theoretical
setting (see for example \cite{Taeuber,Taeuber07})
corresponds to $\langle \phi^2(t,\vec{x}) (\phi
\tilde{\phi})(s,\vec{x}+\vec{r}) \rangle$.

Our findings suggest that for the short-range model the
operator $\phi^2$, corresponding to spin$^2$, 
is quasi-primary and so is the corresponding response field
$\tilde{\phi}^2$ (obtained by locally perturbing the
temperature).
The parameters of these two fields
are related to the fields $\phi$ and $\tilde{\phi}$
in the following way: If $\phi$ has the parameters
$(x,\xi,\mu,\beta)$ then the parameters of $\phi^2$ can be
obtained from these by multiplying each parameter by the
factor $2$. Similarly the parameters of
$\tilde{\phi}^2$ are related to those of
$\tilde{\phi}$.
On the other hand, we see that the composite operator $\phi
\tilde{\phi}$ (defined by a perturbation of the
external field $h_2(t,\vec{x})$) is {\it not} quasi-primary,
and neither is the  energy-density operator
$\epsilon(\vec{x})$, even in the short-range model (that last finding
is not surprising, since we have already seen in section~2 that
$\epsilon(\vec{x})$ is related to the gradient of $\phi$).\footnote{In the 
Landau-Ginzbourg classification of primary scaling operators in the minimal 
models of $2D$ conformal field-theory (Ising, Potts etc.), 
one usually has that $\phi$ and eventually a finite number of normal-ordered 
powers $:\phi:^{\ell}$ are primary.}

We now proceed to the long-range model, where $0<\sigma<2$. 
$\mathcal{R}_{22}$ cannot be brought into the
LSI-predicted form, for the same reason as 
mentioned above for the short-range model, namely by comparing the $t+s$ dependence. The response function
${\cal R}_2^{(T)}$ cannot be brought into the LSI-predicted form either,
since it contains a product of the type 
$ \mathcal{F}^{(\mu,\beta)}(t,s;\vec{r})^2 $.
This again cannot be cast
into the general form ($\ref{spacetime_lsi}$), except for $z = 2$. 
In this exceptional case, the special
properties of a Gaussian integral ensure
that $\mathcal{F}^{(\mu,\beta)}(t,s;\vec{r})^2$ can be
rewritten in the form (\ref{spacetime_lsi}) upon redefinition
of parameters. By a similar analysis we find that ${\cal
R}_{\epsilon2}$ does not have the LSI-predicted form.
We conclude that the operator $\phi^2$ is not
quasi-primary under LSI for the long-range model, 
unlike for the short-range case  $\sigma \geq 2$.  

In a similar way, we also find 
that the response functions of the operator 
$\mathcal{O}_\epsilon$, namely $\mathcal{R}_{\epsilon2}$ and
$\mathcal{R}_{\epsilon \epsilon}$, also do not have
the form (\ref{response_lsi}) and (\ref{spacetime_lsi}).

Summarising, we have seen that in the long-range model the
above composite fields, though made of
quasi-primary fields, are not quasi-primary. For the time being,
the order-parameter $\phi$ and the associate response field $\wit{\phi}$
related to a magnetic perturbation remain the only scaling operators with a 
simple transformation under local scale-transformations. This is distinct from  
the short-range case of $z = 2$. It remains an open 
question in which sense the transformation of, say, $\phi^2$ is distinct from 
the one of $\phi$. On the other hand, the
generalised Bargman superselection rule (which follows from the weaker Galilei-invariance alone) has been confirmed
in all cases, by assigning the following (relative)
'masses' to the fields
\BEQ
\mu_\phi = \mu, \quad \mu_{\mathcal{O}_2} = 2 \mu, \quad
\mu_{\mathcal{O}_\epsilon} = 2 \mu ,
\EEQ
and with negative masses to the corresponding response
fields. This is natural because of the
linear structure of the theory.

\subsection{Correlation functions}

In this section, we compare the LSI-prediction for the correlation 
function of the quasi-primary operator $\phi(t,\vec{x})$ with our
exact result, see (\ref{C-fnR:I},\ref{C-fn:IIa},\ref{C-fn:IIb}). 

The LSI-prediction for the correlation function, for fully disordered initial conditions 
with white noise, is
\cite{Baumann07}:
\BEQ
C^{LSI}(t,s;\vec{r}) = C_{\rm init}^{LSI}(t,s;\vec{r}) +
C_{\rm th}^{LSI}(t,s;\vec{r}) ,
\EEQ
with the `initial' part
\BEA
& & \hspace{-1.5cm}C^{LSI}_{\rm init}(t,s; \vec{r}) =c_0
s^{-b_{init} + 2 \beta /z + d/z}
y^{-b_{init} + \lambda_R/z + \beta/z}
(y-1)^{b_{init} + d/z - 2 \lambda_R /z} \nonumber \\
& & \hspace{-0.5cm} \times \int_{\vec{k}} 
|\vec{k}|^{2 \beta}
\exp\left(-\frac{|\vec{k}|^z}{z^2 \mu \II^{2-z}} (t + s)
\right) e^{\II r \cdot \vec{k}} ,
\EEA
and the `thermal' part
\BEA
& &\hspace{-2.0cm} C^{LSI}_{\rm th}(t,s; \vec{r}) = 2 T
s^{-b_{th} + 2 \beta/z + d/z} y^{2\xi/z} (y-1)^{2(1+a') - 2
\lambda_R/z - 4 \xi/z} \nonumber \\ &  & 
\hspace{-1.5cm} \times
\int_0^1 \!\D\theta\, (y-\theta)^{-2(a'+1) + \lambda_R/z + 2
\xi/z + \beta/z + d/z}
(1-\theta)^{-2(a' +1) + \lambda_R/z + 2 \xi/z + \beta/z +
d/z}
 \\ &  & \hspace{-1.5cm}
 \times \theta^{{4}\tilde{\xi}/z} g\left(\frac{1}{y} 
 \frac{y-\theta}{1-\theta} \right) \int_{\vec{k}}  |\vec{k}|^{2 \beta}
 \exp\left(-\frac{|\vec{k}|^z s (y+1 - 2 \theta)}{z^2 \mu
 \II^{(2-z)}}
 \right) e^{\II \vec{r} \cdot\vec{k}} \nonumber .
 \EEA
Here the function $g(u)$ is not determined by the dynamical 
symmetries and $\xi$ and $\tilde{\xi}$ can be considered as
free parameters. 

In case I, the spin-spin correlation function
(\ref{C-fnR:I}) can be rewritten as 
\BEQ
C(t,s;\vec{r}) =  s^{-\digamma} y^{-\digamma/2} 
\int_{\vec{k}} 
e^{\II \vec{r} \cdot \vec{k}} e^{-B |\vec{k}|^\sigma (t+s)} ,
\EEQ
up to a normalisation constant, with $\alpha$ given by (\ref{exp_alpha}).
In this case ($T < T_c$), the contribution coming from the initial noise is
the relevant one \cite{Bray94}, and therefore we should
compare with the spin-spin correlator $C_{init}^{LSI}(t,s;\vec{r})$. 
Indeed we find
for the choice of parameters as given in (\ref{choice_parameter1}),
(\ref{choice_parameter}) and $b_{init} = 0$ that
$ C_{\rm init}^{LSI}(t,s; \vec{r}) = C(t,s, \vec{r})$,
as it should be. 

In case IIa and IIb, the correlation function, as given in
(\ref{C-fn:IIa}) and (\ref{C-fn:IIb}), can be rewritten in
direct space as follows, using again (\ref{exp_alpha}) and up to
normalisation constant,
\BEQ
C_{\rm th}(t,s;\vec{r}) = 2 T s y^{-\digamma/2} \int_0^1 \!\D\theta
\,\theta^{\digamma} \int_{\vec{k}} e^{-B
|\vec{k}|^\sigma s (y + 1 - 2 \theta)} e^{\II \vec{r} \cdot
\vec{k}} .
\EEQ
For the cases $IIa$ ($T = T_c, \sigma < d < 2
\sigma$) and $IIb$ ($T = T_c, d > 2 \sigma$), in
the LSI-prediction the term coming from the
thermal noise is the relevant one \cite{Bray94,Calabrese05}. If we set 
$g(u)=1$ and, in addition to the given choice of
parameters (\ref{choice_parameter1}) and
(\ref{choice_parameter}), let
\BEQ
b_{th} = \frac{d}{z} -1, \qquad \mbox{and} \qquad
\xi = - \frac{1}{4} z \digamma, \quad \tilde{\xi} =
\frac{1}{4} z \digamma,
\EEQ
we find agreement of the LSI-predicted correlation
function $C_{\rm th}^{LSI}(t,s; \vec{r}) = C(t,s; \vec{r})$.

\section{Conclusion}
We have analysed the kinetics of the spherical model with long-range
interactions when quenched onto or below the critical point
$T_c$. For $T<T_c$ we have reproduced the results of Cannas {\it et al.}
\cite{Cann01} for the order-parameter 
and for $T=T_c$ we have derived exact results for the response 
and correlation function of the order parameter. We also considered, 
for $T\leq T_c$, 
various composite fields and derived their ageing exponents and scaling
functions as listed in section~2. We then have carried out 
a detailed test of local scale-invariance using
our analytical results. For this purpose, the long-range spherical model
offers the useful feature that its dynamical exponent $z=\sigma$ depends
continuously on one of the control parameters. 

We have obtained the following results:
\begin{enumerate}
\item Dynamical scaling holds for various composite fields
for quenches onto or below the critical temperature. The
non-equilibrium exponents are given in table~\ref{table1}
and~\ref{table2}. The scaling functions also have been determined.
\item In the kinetic spherical model with short-ranged interactions
($\sigma>2$ and hence $z=2$), 
apart from  the order-parameter field $\phi$, its square too appears 
to be a quasi-primary scaling operator, as tested through
several two-time response and correlation functions.
\item In the long-range spherical model, the first tests of the space-time
response in a system with a tunable dynamical exponent have been performed. 
This shows that the formulation of LSI with $z\ne 2$, which we proposed 
earlier \cite{Henkel02}, even with the recent improvements given in 
\cite{Roethlein06},
does not describe the exact result for ${\cal R}_{11}$ when $0<z<2$, 
although that formulation did pass previous tests when $z=2$ \cite{Henkel03} 
or $z=4$ \cite{Roethlein06,Baumann07b}. 
\item As can be seen from the fractional differential equation satisfied
by the space-time response function, the precise definition of the fractional
derivative used is crucial. We shall present elsewhere a systematic
construction of new generators of local scale-invariance
\cite{Baumann07} where we shall also show that {\em all} previous tests where
$z=2$ or $z=4$ are passed by the new formulation. Here we have seen that the
exact results from the long-range spherical model are completely consistent
with the new formulation of local scale-invariance. 
\item In contrast to the short-range case where $z=2$, the spin-squared 
field in the long-range model is no longer
described by a quasi-primary scaling operator. This calls for a more systematic
analysis, since it indicates that there might be new ways, not readily 
realized in conformal invariance, of non-quasi-primary scaling operators. 
\item Both the two-time response and the correlation function of the
order-parameter field $\phi$ are fully compatible with local scale-invariance
in the entire range $0<z=\sigma<2$. 
\end{enumerate}
While the analytical results presented here certainly 
provide useful information,
the eventual confirmation of local scale-invariance might appear fairly 
natural since the underlying Langevin equation is {\em linear}. Indeed, for 
linear Langevin equations there is a direct proof of local 
scale-invariance which uses a decomposition of the Langevin equation 
into a `deterministic part' for which non-trivial local scale-symmetries can 
be mathematically proven and a `noise part' 
\cite{Picone04,Henkel07c,Baumann07}. For non-linear Langevin equations
the formal proof of non-trivial symmetries of the `deterministic part' is 
still difficult, although progress has been made \cite{Stoimenov05}. 
In the absence of exact solutions for models described in 
terms of non-linear Langevin equations numerical tests going beyond merely
checking the autoresponse function $R(t,s;\vec{0})$ will be
required and it will be useful to be able to vary the value of the
dynamical exponent $z$. In this context, a natural candidate for such
studies is the disordered Ising model quenched to $T<T_c$, where it is
already known that $z$ depends continuously on the disorder and on temperature,
see \cite{Paul04,Henkel06b,Henkel07b} and references therein. Furthermore,
its Langevin equation is non-linear. We hope to be able soon to report tests
on the space-time behaviour of response and correlators in this model which
should provide useful information on whether LSI with $z\ne 2$ can really
be extended beyond the simple solvable systems studied so far.   


\annexe{Correlations and responses}

We present briefly the calculational details that lead to the expressions
given in section~\ref{corr-resp}. In evaluating the expectation value of the 
composite operators we use Wick's contraction, which is applicable in our model
if, apart from the noise, the initial spin distribution for 
$S(0,\vec{x})$ is also 
Gaussian \cite{Zinn02}. By Wick's contraction and Fourier transforming, we get
\begin{eqnarray}
\label{s2-en-corr1}
{\cal C}_{2 \epsilon}(t,t';\vec{x} - \vec{x}') = 
\int_{{\Lambda}_{\vec{k}}, {\Lambda}_{\vec{k}'}}
e^{\II (\vec{k}+\vec{k}')\cdot(\vec{x}-\vec{x}')}
\left( \omega_{\vec{k}} + \omega_{\vec{k}'} \right)
C(t,t';\vec{k})C(t,t';\vec{k}')  \nonumber \\
= - \frac{1}{g(t)} \partial_t 
\int_{{\Lambda}_{\vec{k}}, {\Lambda}_{\vec{k}'}}
e^{\II (\vec{k}+\vec{k}')\cdot(\vec{x}-\vec{x}')}
g(t) C(t,t';\vec{k})C(t,t';\vec{k}') ,
\end{eqnarray}
resulting in equation (\ref{s2-en-corr}). Similarly,
\begin{eqnarray}
\label{en-en-corr1}
{\cal C}_{\epsilon \epsilon}(t,t'; \vec{x} - \vec{x}') = 
\frac{1}{2} \int_{{\Lambda}_{\vec{k}}, {\Lambda}_{\vec{k}'}}
e^{\II (\vec{k}+\vec{k}')\cdot(\vec{x}-\vec{x}')}
\left( \omega_{\vec{k}} + \omega_{\vec{k}'} \right)^2 
C(t,t';\vec{k})C(t,t';\vec{k}')  \nonumber \\
= \frac{1}{2g(t)} \partial_t^2 
\int_{{\Lambda}_{\vec{k}}, {\Lambda}_{\vec{k}'}}
e^{\II (\vec{k}+\vec{k}')\cdot(\vec{x}-\vec{x}')} 
g(t) C(t,t';\vec{k})C(t,t';\vec{k}') ,
\end{eqnarray}
gives the equation (\ref{en-en-corr}).

When 
${\cal H} \rightarrow {\cal H} - \sum_{t,\vec{x}} h_1(t,\vec{x}) S(t,\vec{x})$,
the solution to the corresponding Langevin equation is  
\BEQ
\label{h1-soln}
\wht{S}_1(t,\vec{k}; h_1)  = \wht{S}(t,\vec{k})
+ \frac {e^{-\omega_{\vec{k}} t} } {\sqrt{g(t)}}
\int_0^t \!\D\tau\, e^ {\omega_{\vec{k}} \tau } \sqrt{g(\tau)} 
h_1(\tau,\vec{k}),
\EEQ 
where the first term on the right-hand side is the 
unperturbed solution as given in
equation (\ref{solution}). Differentiating the above expression 
with respect to 
$h_1(t',\vec{x}')$ and Fourier transforming back gives 
${\cal R}_{11}(t,t';\vec{x} - \vec{x}')$, while both
${\cal R}_{21}(t,t';\vec{x} - \vec{x}')$ and  
${\cal R}_{\epsilon1}(t,t';\vec{x} - \vec{x}')$ vanish since
$\langle \wht{S}(t,\vec{k})\rangle = 0$. 
   
When 
${\cal H} \rightarrow {\cal H} - \sum_{t,\vec{x}} h_2(t,\vec{x}) 
S^2(t,\vec{x})$, 
the solution to the corresponding Langevin equation is  
\BEQ
\label{h2-soln}
\wht{S}_2(t,\vec{k};h_2)  = \wht{S}(t,\vec{k}) +
2\int_0^t\!\!\!\D\tau\, e^ {\omega_{\vec{k}} (\tau-t) } 
\sqrt{\frac{g(\tau)}{g(t)}} 
\int_{{\Lambda}_{\vec{k}'}} \!\!\!
\wht{S}_2(\tau,\vec{k'}; h_2) h_2(\tau,\vec{k}-\vec{k'})  .
\EEQ 
Here $\wht{S}_2(\tau,\vec{k'}; h_2)$ on the right-hand 
side can be replaced by
$\wht{S}(\tau,\vec{k'})$, while evaluating the response functions, 
since the difference is of order $O(h_2^2)$. 
Clearly ${\cal R}_{12}(t,t';\vec{x} - \vec{x}')$ vanishes 
since the initial magnetisation is zero. Using the above equation we get,
\begin{eqnarray}
\label{h2-soln:1}
\left\langle \wht{S}(t,\vec{k}) \frac{\delta \wht{S}_2(t,\vec{k'};h_2) } 
{\delta h_2(t',\vec{x}')} \right\rangle_{h_2 =0} 
=  2 \sqrt{\frac{g(t')}{g(t)}}  e^{-\omega_{\vec{k}'}(t-t')}  
 C(t,t';\vec{k})   e^{-\II(\vec{k}+\vec{k}')\cdot\vec{x}'},  
\end{eqnarray}
and then multiplying it by $2\exp(\II(\vec{k}+\vec{k}')\cdot\vec{x})$ and 
integrating over ${\Lambda}_{\vec{k}}$ and ${\Lambda}_{\vec{k}'}$ results in 
equation (\ref{22-resp}). When we multiply equation (\ref{h2-soln:1}) by
$2 \omega_{\vec{k}} \exp(\II(\vec{k}+\vec{k}')\cdot\vec{x})$ 
and integrate over ${\Lambda}_{\vec{k}}$ and ${\Lambda}_{\vec{k}'}$, we get
\BEQ
\label{en2-resp1}
{\cal R}_{\epsilon 2}(t,t';\vec{x} - \vec{x}') = 
4 \sqrt{\frac{g(t')}{g(t)}}  C(t,t';\vec{x} - \vec{x}')
\int_{{\Lambda}_{\vec{k}}} \omega_{\vec{k}} 
e^{\II \vec{k}\cdot(\vec{x}- \vec{x}')} e^{-\omega_{\vec{k}}(t-t')} ,
\EEQ
which is rewritten as in equation (\ref{en2-resp}).

When ${\cal H} \rightarrow 
{\cal H} - \sum_{t,\vec{x}} h_{\epsilon}(t,\vec{x}) 
{\cal O}_{\epsilon}(t,\vec{x})$,
the solution to the corresponding Langevin equation   
$\wht{S}_{\epsilon}(t,\vec{k};\epsilon) =
\wht{S}(t,\vec{k}) + \delta \wht{S}_{\epsilon}(t,\vec{k};\epsilon) $, where
\BEQ
\label{en-soln}
\delta \wht{S}_{\epsilon}(t,\vec{k};\epsilon)  =
\int_0^t\!\!\D\tau\, e^ {\omega_{\vec{k}} (\tau-t) } \sqrt{\frac{g(\tau)}{g(t)}} 
\int_{{\Lambda}_{\vec{k}'}} \!\!
\left( \omega_{\vec{k}} + \omega_{\vec{k}'} \right)
\wht{S}_{\epsilon}(\tau,\vec{k}';\epsilon)  h_{\epsilon}(\tau,\vec{k}-\vec{k'})  .
\EEQ 
{}From the above equation we get
\begin{eqnarray}
\label{2en-resp:1}
\left\langle \wht{S}(t,\vec{k})
\frac{\delta  \wht{S}_{\epsilon}(t,\vec{k}';\epsilon)  } 
{\delta h_{\epsilon}(t',\vec{x}')}  
\right\rangle_{\epsilon =0} 
=  \sqrt{\frac{g(t')}{g(t)}}  e^{\omega_{\vec{k}'}(t'-t)} 
e^{-\II(\vec{k}+\vec{k}')\cdot\vec{x}'}  
\left( \omega_{\vec{k}} + \omega_{\vec{k}'} \right) C(t,t';\vec{k}) 
\nonumber \\
=  -\frac{\sqrt{g(t')}}{g(t)}  \partial_t 
\left( e^{-\omega_{\vec{k}'}(t-t')} e^{-\II(\vec{k}+\vec{k}')\cdot\vec{x}'}  
\sqrt{g(t)} C(t,t';\vec{k}) \right).
\end{eqnarray}
Now multiply this equation by $2\exp(i(\vec{k}+\vec{k}')\cdot\vec{x})$ and 
integrate over ${\Lambda}_{\vec{k}}$ and ${\Lambda}_{\vec{k}'}$ 
to get equation (\ref{2en-resp}). If we 
multiply the equation (\ref{2en-resp:1}) by a factor 
$(\omega_{\vec{k}} + \omega_{\vec{k}'} )$ 
then the only change in the last term is that $\partial_t$ gets replaced by 
$-\partial_t^2 $. Further, multiplying by 
$\exp(i(\vec{k}+\vec{k}')\cdot\vec{x})$ and 
integrating over ${\Lambda}_{\vec{k}}$ and ${\Lambda}_{\vec{k}'}$
results in equation (\ref{enen-resp}).

When the temperature $T \rightarrow T + T'(t,\vec{x})$ is shifted then 
the solution $\wht{S}_T(t,\vec{k};T')$ evolves just as given in 
equation (\ref{solution}), 
where the mean is $\langle \wht{\eta}(t,\vec{k})\rangle_{T'}= 0$,
but the variance becomes
\BEQ
\label{noise-corr:T'}
\left\langle \wht{\eta}(t,\vec{k})\wht{\eta}(t',\vec{k}')\right\rangle_{T'}=
\left\langle \wht{\eta}(t,\vec{k})\wht{\eta}(t',\vec{k}')\right\rangle  +
2 T'_{\vec{k}+\vec{k}'}(t)  \delta(t-t').
\EEQ
This implies $\langle \wht{S}_T(t,\vec{k};T') \rangle$ is independent of $T'$ 
and
\BEQ
\label{corr:T'}
\left. \frac{\delta} {\delta{T'(t',\vec{x}')} } 
\langle \wht{S}(t,\vec{k})\wht{S}(t,\vec{k}')\rangle_{T'} \right|_{T'=0} =
2 \frac{g(t')}{g(t)} e^{-(\omega_{\vec{k}} + \omega_{\vec{k}'})(t-t')}
e^{-\II(\vec{k}+\vec{k}')\cdot\vec{x}'} 
\EEQ
Now multiplying this equation by $\exp(\II(\vec{k}+\vec{k}')\cdot\vec{x})$,
and then integrating over $\Lambda_{\vec{k}}$ and $\Lambda_{\vec{k}'}$ 
results in equation (\ref{2T-resp}). Similarly,
multiplying by $\exp(\II(\vec{k}+\vec{k}')\cdot\vec{x})$ along with the factor 
$(\omega_{\vec{k}} + \omega_{\vec{k}'} )/2$, and then  
integrating over $\Lambda_{\vec{k}}$ and $\Lambda_{\vec{k}'}$ results in 
equations (\ref{enT-resp}).

\noindent 
{\bf Acknowledgements:}
FB acknowledges the support by the Deutsche Forschungsgemeinschaft
through grant no. PL 323/2. 
FB and MH are supported by the franco-german binational
programme PROCOPE.
SD acknowledges the generous support by the people of South Korea.


\end{document}